\documentclass[10pt,a4paper,onecolumn]{article}
\usepackage{marginnote}
\usepackage{graphicx}
\usepackage[rgb,svgnames]{xcolor}
\usepackage{authblk,etoolbox}
\usepackage{titlesec}
\usepackage{calc}
\usepackage{tikz}
\usepackage[pdfa]{hyperref}
\usepackage{hyperxmp}
\hypersetup{%
    unicode=true,
    pdfapart=3,
    pdfaconformance=B,
    pdftitle={GWSurrogate: A Python package for gravitational wave
surrogate models},
    pdfauthor={Scott E. Field, Vijay Varma, Jonathan Blackman, Bhooshan
Gadre, Chad R. Galley, Tousif Islam, Keefe Mitman, Michael Pürrer,
Adhrit Ravichandran, Mark A. Scheel, Leo C. Stein, Jooheon Yoo},
    pdfpublication={Journal of Open Source Software},
    pdfpublisher={Open Journals},
    pdfissn={2475-9066},
    pdfpubtype={journal},
    pdfvolumenum={},
    pdfissuenum={},
    pdfdoi={10.21105/joss.07073},
    pdfcopyright={Copyright (c) 2025, Scott E. Field, Vijay Varma,
Jonathan Blackman, Bhooshan Gadre, Chad R. Galley, Tousif Islam, Keefe
Mitman, Michael Pürrer, Adhrit Ravichandran, Mark A. Scheel, Leo C.
Stein, Jooheon Yoo},
    pdflicenseurl={http://creativecommons.org/licenses/by/4.0/},
    colorlinks=true,
    linkcolor=[rgb]{0.0, 0.5, 1.0},
    citecolor=Blue,
    urlcolor=[rgb]{0.0, 0.5, 1.0},
    breaklinks=true
}
\immediate\pdfobj stream attr{/N 3} file{sRGB.icc}
\pdfcatalog{%
  /OutputIntents [
    <<
      /Type /OutputIntent
      /S /GTS_PDFA1
      /DestOutputProfile \the\pdflastobj\space 0 R
      /OutputConditionIdentifier (sRGB)
      /Info (sRGB)
    >>
  ]
}
\usepackage{caption}
\usepackage{orcidlink}
\usepackage{tcolorbox}
\usepackage{amssymb,amsmath}
\usepackage{ifxetex,ifluatex}
\usepackage{seqsplit}
\usepackage{xstring}

\usepackage{float}
\let\origfigure\figure
\let\endorigfigure\endfigure
\renewenvironment{figure}[1][2] {
    \expandafter\origfigure\expandafter[H]
} {
    \endorigfigure
}

\usepackage{fixltx2e} 

\NewDocumentCommand\citeproctext{}{}
\NewDocumentCommand\citeproc{mm}{%
  \begingroup\def\citeproctext{#2}\cite{#1}\endgroup}
\makeatletter
 \let\@cite@ofmt\@firstofone
 \def\@biblabel#1{}
 \def\@cite#1#2{{#1\if@tempswa , #2\fi}}
\makeatother
\newlength{\cslhangindent}
\newlength{\csllabelwidth}
\newenvironment{CSLReferences}[2] 
 {\begin{list}{}{%
  \setlength{\itemindent}{0pt}
  \setlength{\leftmargin}{0pt}
  \setlength{\parsep}{0pt}
  \ifodd #1
   \setlength{\leftmargin}{\cslhangindent}
   \setlength{\itemindent}{-1\cslhangindent}
  \fi
  \setlength{\itemsep}{#2\baselineskip}}}
 {\end{list}}
\usepackage{calc}

\usepackage[top=3.5cm, bottom=3cm, right=1.5cm, left=1.0cm,
            headheight=2.2cm, reversemp, includemp, marginparwidth=4.5cm]{geometry}

\titleformat{\section}
  {\normalfont\sffamily\Large\bfseries}
  {}{0pt}{}
\titleformat{\subsection}
  {\normalfont\sffamily\large\bfseries}
  {}{0pt}{}
\titleformat{\subsubsection}
  {\normalfont\sffamily\bfseries}
  {}{0pt}{}
\titleformat*{\paragraph}
  {\sffamily\normalsize}

\usepackage{fancyhdr}
\makeatletter
\let\ps@plain\ps@fancy
\fancyheadoffset[L]{4.5cm}
\fancyfootoffset[L]{4.5cm}

\definecolor{linky}{rgb}{0.0, 0.5, 1.0}

\newtcolorbox{repobox}
   {colback=red, colframe=red!75!black,
     boxrule=0.5pt, arc=2pt, left=6pt, right=6pt, top=3pt, bottom=3pt}

\newcommand{\ExternalLink}{%
   \tikz[x=1.2ex, y=1.2ex, baseline=-0.05ex]{%
       \begin{scope}[x=1ex, y=1ex]
           \clip (-0.1,-0.1)
               --++ (-0, 1.2)
               --++ (0.6, 0)
               --++ (0, -0.6)
               --++ (0.6, 0)
               --++ (0, -1);
           \path[draw,
               line width = 0.5,
               rounded corners=0.5]
               (0,0) rectangle (1,1);
       \end{scope}
       \path[draw, line width = 0.5] (0.5, 0.5)
           -- (1, 1);
       \path[draw, line width = 0.5] (0.6, 1)
           -- (1, 1) -- (1, 0.6);
       }
   }

\definecolor{c53baa1}{RGB}{83,186,161}
\definecolor{c202826}{RGB}{32,40,38}

\patchcmd{\@maketitle}{center}{flushleft}{}{}
\patchcmd{\@maketitle}{center}{flushleft}{}{}
\patchcmd{\@maketitle}{\LARGE}{\LARGE\sffamily}{}{}
\def\maketitle{{%
  
  \AB@maketitle}}
\renewcommand\AB@affilsepx{ \protect\Affilfont}
\renewcommand\AB@affilnote[1]{{\bfseries #1}\hspace{3pt}}
\renewcommand{\affil}[2][]%
   {\newaffiltrue\let\AB@blk@and\AB@pand
      \if\relax#1\relax\def\AB@note{\AB@thenote}\else\def\AB@note{#1}%
        \setcounter{Maxaffil}{0}\fi
        \begingroup
        \let\href=\href@Orig
        \let\protect\@unexpandable@protect
        \def\thanks{\protect\thanks}\def\footnote{\protect\footnote}%
        \@temptokena=\expandafter{\AB@authors}%
        {\def\\{\protect\\\protect\Affilfont}\xdef\AB@temp{#2}}%
         \xdef\AB@authors{\the\@temptokena\AB@las\AB@au@str
         \protect\\[\affilsep]\protect\Affilfont\AB@temp}%
         \gdef\AB@las{}\gdef\AB@au@str{}%
        {\def\\{, \ignorespaces}\xdef\AB@temp{#2}}%
        \@temptokena=\expandafter{\AB@affillist}%
        \xdef\AB@affillist{\the\@temptokena \AB@affilsep
          \AB@affilnote{\AB@note}\protect\Affilfont\AB@temp}%
      \endgroup
       \let\AB@affilsep\AB@affilsepx
}
\makeatother

\renewcommand\Affilfont{\sffamily\small\mdseries}
\ifnum 0\ifxetex 1\fi\ifluatex 1\fi=0 
  \usepackage[T1]{fontenc}
  \usepackage[utf8]{inputenc}

\else 
  \ifxetex
    \usepackage{mathspec}
    \usepackage{fontspec}

  \else
    \usepackage{fontspec}
  \fi
  \defaultfontfeatures{Scale=MatchLowercase}
  \defaultfontfeatures[\sffamily]{Ligatures=TeX}
  \defaultfontfeatures[\rmfamily]{Ligatures=TeX,Scale=1}

\fi
\IfFileExists{upquote.sty}{\usepackage{upquote}}{}
\IfFileExists{microtype.sty}{%
\usepackage{microtype}
\UseMicrotypeSet[protrusion]{basicmath} 
}{}

\PassOptionsToPackage{usenames,dvipsnames}{color} 
\ifLuaTeX
\usepackage[bidi=basic]{babel}
\else
\usepackage[bidi=default]{babel}
\fi
\babelprovide[main,import]{american}

\def\languageshorthands#1{}

\usepackage{graphicx,grffile}
\makeatletter
\def\maxwidth{\ifdim\Gin@nat@width>\linewidth\linewidth\else\Gin@nat@width\fi}
\def\maxheight{\ifdim\Gin@nat@height>\textheight\textheight\else\Gin@nat@height\fi}
\makeatother
\setkeys{Gin}{width=\maxwidth,height=\maxheight,keepaspectratio}
\IfFileExists{parskip.sty}{%
\usepackage{parskip}
}{
\setlength{\parindent}{0pt}
\setlength{\parskip}{6pt plus 2pt minus 1pt}
}
\providecommand{\tightlist}{%
  \setlength{\itemsep}{0pt}\setlength{\parskip}{0pt}}
\ifx\paragraph\undefined\else
\let\oldparagraph\paragraph
\renewcommand{\paragraph}[1]{\oldparagraph{#1}\mbox{}}
\fi
\ifx\subparagraph\undefined\else
\let\oldsubparagraph\subparagraph
\renewcommand{\subparagraph}[1]{\oldsubparagraph{#1}\mbox{}}
\fi
\ifLuaTeX
  \usepackage{selnolig}  
\fi

\title{GWSurrogate: A Python package for gravitational wave surrogate
models}

\author[1,5%
*%
]{Scott E. Field%
  \,\orcidlink{0000-0002-6037-3277}\,%
}
\author[1%
*%
]{Vijay Varma%
  \,\orcidlink{0000-0002-9994-1761}\,%
}
\author[2%
]{Jonathan Blackman%
}
\author[3%
]{Bhooshan Gadre%
  \,\orcidlink{0000-0002-1534-9761}\,%
}
\author[2%
]{Chad R. Galley%
}
\author[1,4%
]{Tousif Islam%
  \,\orcidlink{0000-0002-3434-0084}\,%
}
\author[2%
]{Keefe Mitman%
  \,\orcidlink{0000-0003-0276-3856}\,%
}
\author[5%
]{Michael Pürrer%
  \,\orcidlink{0000-0002-3329-9788}\,%
}
\author[1%
]{Adhrit Ravichandran%
}
\author[2%
]{Mark A. Scheel%
  \,\orcidlink{0000-0001-6656-9134}\,%
}
\author[6%
]{Leo C. Stein%
  \,\orcidlink{0000-0001-7559-9597}\,%
}
\author[7%
]{Jooheon Yoo%
  \,\orcidlink{0000-0002-3251-0924}\,%
}

\affil[1]{Department of Mathematics and Center for Scientific Computing
\& Data Science Research, University of Massachusetts, Dartmouth, MA
02747, USA%
}
\affil[2]{Theoretical Astrophysics 350-17, California Institute of
Technology, Pasadena, California 91125, USA%
}
\affil[3]{Institute for Gravitational and Subatomic Physics (GRASP),
Utrecht University, 3584 CC Utrecht, The Netherlands%
}
\affil[4]{Kavli Institute for Theoretical Physics, University of
California Santa Barbara, CA 93106, USA%
}
\affil[5]{Department of Physics and Center for Computational Research,
East Hall, University of Rhode Island, Kingston, RI 02881, USA%
}
\affil[6]{Department of Physics and Astronomy, The University of
Mississippi, University, MS 38677, USA%
}
\affil[7]{Cornell Center for Astrophysics and Planetary Science, Cornell
University, Ithaca, New York 14853, USA%
}
\affil[*]{These authors contributed equally.}
\date{\vspace{-2.5ex}}

\begin{document}
\maketitle

\marginpar{

  \begin{flushleft}
  \sffamily\small

  {\bfseries DOI:} \href{https://doi.org/10.21105/joss.07073}{\color{linky}{10.21105/joss.07073}}

  \vspace{2mm}
    {\bfseries Software}
  \begin{itemize}
    \setlength\itemsep{0em}
    \item \href{https://github.com/openjournals/joss-reviews/issues/7073}{\color{linky}{Review}} \ExternalLink
    \item \href{https://github.com/sxs-collaboration/gwsurrogate}{\color{linky}{Repository}} \ExternalLink
    \item \href{https://doi.org/10.5281/zenodo.14933936}{\color{linky}{Archive}} \ExternalLink
  \end{itemize}

  \vspace{2mm}
  
    \par\noindent\hrulefill\par

  \vspace{2mm}

  {\bfseries Editor:} \href{https://plaplant.github.io/}{Paul La
Plante} \ExternalLink
  \,\orcidlink{0000-0002-4693-0102} \\
  \vspace{1mm}
    {\bfseries Reviewers:}
  \begin{itemize}
  \setlength\itemsep{0em}
    \item \href{https://github.com/GarethCabournDavies}{@GarethCabournDavies}
    \item \href{https://github.com/Ceciliogq}{@Ceciliogq}
    \end{itemize}
    \vspace{2mm}
  
    {\bfseries Submitted:} 03 June 2024\\
    {\bfseries Published:} 29 March 2025

  \vspace{2mm}
  {\bfseries License}\\
  Authors of papers retain copyright and release the work under a Creative Commons Attribution 4.0 International License (\href{https://creativecommons.org/licenses/by/4.0/}{\color{linky}{CC BY 4.0}}).

  \end{flushleft}
}

\section{Summary}\label{summary}

Gravitational waves are ripples in space-time caused by the motion of
massive objects. One of the most astrophysically important sources of
gravitational radiation is caused by two orbiting compact objects, such
as black holes and neutron stars, that slowly inspiral and merge. The
motion of these massive objects generates gravitational waves that
radiate to the far field where gravitational-wave detectors can observe
them. Complicated partial or ordinary differential equations govern the
entire process.

Traditionally, the dynamics of compact binary systems and the emitted
gravitational waves have been computed by expensive simulation codes
that can take days to months to run. A key simulation output is the
gravitational wave signal for a particular set of parameter values
describing the system, such as the black holes' masses and spins. The
computed signal is required for a diverse range of multiple-query
applications, such as template bank generation for searches, parameter
estimation, mock data analysis, studies of model bias, and tests of
general relativity, to name a few. In such settings, the high-fidelity
signal computed from differential equations is often too slow to be
directly used.

\begin{figure}
\centering
\includegraphics{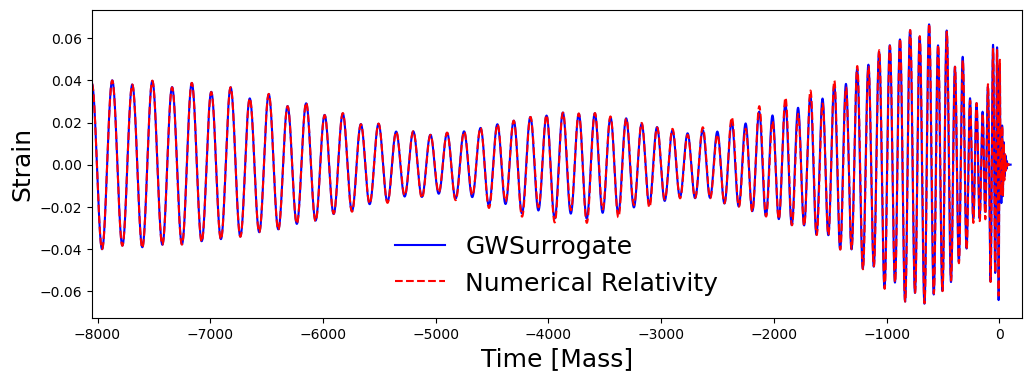}
\caption{Example gravitational wave prediction from a surrogate model
compared with numerical relativity for a precessing binary black hole
system. This particular numerical relativity simulation took 70,881
CPU-hours (about 1.75 months using 56 cores on the supercomputer
Frontera), while the surrogate model can be evaluated in about 100
milliseconds. \label{fig:gws}}
\end{figure}

Surrogate models offer a practical way to dramatically accelerate model
evaluation while retaining the high-fidelity accuracy of the expensive
simulation code; an example is shown in Fig. \ref{fig:gws}. Surrogate
models can be constructed in various ways, but what separates these
models from other modeling frameworks is that they are primarily
data-driven. Given a training set of gravitational waveform data
sampling the parameter space, a model is built by following three steps:

\begin{enumerate}
\def\labelenumi{\arabic{enumi}.}
\tightlist
\item
  Feature extraction: the waveform is decomposed into \emph{data pieces}
  that are simple to model,
\item
  Dimensionality reduction: each data piece is approximated by a
  low-dimensional vector space, which reduces the degrees of freedom we
  need to model, and
\item
  Regression: fitting and regression techniques are applied to the
  low-dimensional representation of each data piece over the parameter
  space defining the model.
\end{enumerate}

These model-building steps result in an HDF5 file defining the surrogate
model's data and structure, which is stored on Zenodo. The GWSurrogate
package provides access to these models through its catalog interface,
and all available models and their associated HDF5 files can be found in
\texttt{gwsurrogate.catalog.list()}. For a recent overview of surrogate
modeling as used in gravitational wave astrophysics, please see Section
5 of Afshordi et al.
(\citeproc{ref-LISAConsortiumWaveformWorkingGroup:2023arg}{2023}).

The development of GWSurrogate is hosted on
\href{https://github.com/sxs-collaboration/gwsurrogate}{GitHub} and
distributed through both
\href{https://pypi.org/project/gwsurrogate/}{PyPI} and
\href{https://anaconda.org/conda-forge/gwsurrogate/}{Conda}. Quick start
guides are found on the project's
\href{https://github.com/sxs-collaboration/gwsurrogate}{homepage} while
model-specific documentation is described through a collection of
model-specific
\href{https://github.com/sxs-collaboration/gwsurrogate/tree/master/tutorial}{Jupyter
notebooks}. Automated testing is run on
\href{https://github.com/sxs-collaboration/gwsurrogate/actions}{GitHub
Actions}.

\section{Statement of need}\label{statement-of-need}

GWSurrogate is a Python package first introduced in 2013 to provide an
intuitive interface for working with gravitational wave surrogate
models. Specifically, GWSurrogate gravitational wave models provide
evaluation of \[
 h_{\tt S}(t, \theta, \phi;\Lambda) = \sum^{\infty}_{\ell=2} \sum_{m=-\ell}^{\ell} h_{\tt S}^{\ell m}(t;\Lambda) ~^{-2}Y_{\ell m}(\theta, \phi) \,,
\] where \(^{-2}Y_{\ell m}\) are the spin\(=-2\) weighted spherical
harmonics and \(\Lambda\) describes the model's parameterization. The
surrogate model provides fast evaluations for the modes,
\(h_{\tt S}^{\ell m}\). As described more fully in the documentation,
the high-level API allows users direct access to the modes
\(\{h_{\tt S}^{\ell m}(t)\}\) (as a Python dictionary) or assembles the
sum \(h_{\tt S}(t, \theta, \phi)\) at a particular location
\((\theta, \phi)\). The models implemented in GWSurrogate are intended
to be used in production data analysis efforts. As such,

\begin{itemize}
\tightlist
\item
  computationally expensive operations (e.g., interpolation onto uniform
  time grids) are implemented by wrapping low-level C code for speed,
  whereas GWSurrogate provides a user-friendly interface to the
  high-level waveform evaluation API,
\item
  models implemented in GWSurrogate follow the waveform convention
  choices of the LIGO-Virgo-Kagra collaboration, thus ensuring that
  downstream data analysis codes can use GWSurrogate models without
  needing to worry about different conventions, and
\item
  GWSurrogate models can be directly evaluated in either physical units
  (often used in data analysis studies) and dimensionless units (often
  used in theoretical studies) where all dimensioned quantities are
  expressed in terms of the system's total mass.
\end{itemize}

Currently, there are 16 supported surrogate models
(\citeproc{ref-Barkett:2019tus}{Barkett et al., 2020};
\citeproc{ref-Blackman:2015pia}{Blackman et al., 2015},
\citeproc{ref-Blackman:2017dfb}{2017};
\citeproc{ref-Field:2013cfa}{Field et al., 2014};
\citeproc{ref-Gadre:2022sed}{Gadre et al., 2024};
\citeproc{ref-Islam:2022laz}{Islam et al., 2022};
\citeproc{ref-OShaughnessy:2017tak}{O'Shaughnessy et al., 2017};
\citeproc{ref-Rifat:2019ltp}{Rifat et al., 2020};
\citeproc{ref-Varma:2018mmi}{Varma, Field, Scheel, Blackman, Kidder, et
al., 2019}; \citeproc{ref-Varma:2019csw}{Varma, Field, Scheel, Blackman,
Gerosa, et al., 2019}; \citeproc{ref-Yoo:2022erv}{Yoo et al., 2022},
\citeproc{ref-Yoo:2023spi}{2023}), with additional models under
development (\citeproc{ref-Islam:2021mha}{Islam et al., 2021};
\citeproc{ref-Rink:2024swg}{Rink et al., 2024}). These models vary in
their duration, included physical effects (e.g.~nonlinear memory, tidal
forces, harmonic modes retained, eccentricity, mass ratio extent,
precession effects, etc), and underlying solution method (e.g.~Effective
One Body, numerical relativity, and black hole perturbation theory).
Details about all models can be found by doing
\texttt{gwsurrogate.catalog.list(verbose=True)}, while the GWSurrogate
\href{https://github.com/sxs-collaboration/gwsurrogate}{homepage}
summarizes the state-of-the-art models for each particular problem.
Certain models allow for additional functionality such as returning the
dynamics of the binary black hole. These special features are described
further in model-specific
\href{https://github.com/sxs-collaboration/gwsurrogate/tree/master/tutorial}{example
notebooks}.

Several other software packages are available for waveform generation,
including tools for effective-one-body models
(\citeproc{ref-Mihaylov:2023bkc}{Mihaylov et al., 2023};
\citeproc{ref-Nagar:2020pcj}{Nagar et al., 2020}), ringdown signals
(\citeproc{ref-pyRing}{Carullo et al., 2023};
\citeproc{ref-Isi:2021iql}{Isi \& Farr, 2021};
\citeproc{ref-alex_nitz_2024_10473621}{Nitz et al., 2024}),
extreme-mass-ratio inspiral systems through the Black Hole Perturbation
Toolkit's \texttt{FastEMRIWaveforms} and \texttt{BHPTNRSurrogate}
packages (\citeproc{ref-BHPToolkit}{\emph{{Black Hole Perturbation
Toolkit}}, n.d.}), and the \texttt{Ripple} framework that enables
specialized acceleration techniques using JAX
(\citeproc{ref-Edwards:2023sak}{Edwards et al., 2024}). Among these,
LALSuite (\citeproc{ref-lalsuite}{LIGO Scientific Collaboration et al.,
2018}) stands out as the most comprehensive, offering the largest
collection of waveform models via its LALSimulation subpackage, which
includes Python bindings and the new Python-based gwsignal waveform
generator. While GWSurrogate shares similarities with LALSuite in
providing a variety of models, it differs by exclusively focusing on
surrogate models. Notably, GWSurrogate includes many state-of-the-art
numerical relativity models that are only available through its library,
whereas LALSuite offers a broader but less specialized collection.

\section{Acknowledgements}\label{acknowledgements}

We acknowledge our many close collaborators for their contribution to
the development of surrogate models. We further acknowledge the
community of GWSurrogate users who have contributed pull requests and
opened issues, including Kevin Barkett, Mike Boyle, Collin Capano, Dwyer
Deighan, Raffi Enficiaud, Oliver Jennrich, Gaurav Khanna, Duncan
Macleod, Alex Nitz, Seth Olsen, Swati Singh, and Avi Vajpeyi.
GWSurrogate has been developed over the past 10 years with continued
support from the National Science Foundation, most recently through NSF
grants PHY-2110496, PHY-2309301, DMS-2309609, and AST-2407454. This work
was partly supported by UMass Dartmouth's Marine and Undersea Technology
(MUST) research program funded by the Office of Naval Research (ONR)
under grant no. N00014-23-1-2141. This work was also supported in part
by the Sherman Fairchild Foundation, by NSF Grants PHY-2207342 and
OAC-2209655 at Cornell, and by NSF Grants PHY-2309211, PHY-2309231, and
OAC-2209656 at Caltech.

\section*{References}\label{references}
\addcontentsline{toc}{section}{References}

\phantomsection\label{refs}

\end{document}